\documentclass{ws-procs961x669}            % default, citations in superscript
\begin{document}
\title{The internal rotation of low-mass stars from solar
and stellar seismology}

\author{G. Buldgen$^*$ and P. Eggenberger}

\address{Astronomy Department, University of Geneva,\\
Versoix, 1290, Switzerland\\
$^*$E-mail: Gael.Buldgen@unige.ch}

\begin{abstract}
The possibility of measuring the internal rotation of the Sun and stars thanks to helio- and asteroseismology offers tremendous constraints on hydro- and magnetohydrodynamical processes acting in stellar interiors. Understanding the processes responsible for the transport of angular momentum in stellar interiors is crucial as they will also influence the transport of chemicals and thus the evolution of stars. Here we present some of the key results obtained in both fields and how detailed seismic analyses can provide stringent constraints on the physics of angular momentum transport in the interior of low mass stars and potentially rule out some candidates. 
\end{abstract}


\bodymatter

\section{Introduction}\label{sec:Intro}

Low mass stars ($\approx 1 M_{\odot}$) are mostly characterized by the presence in their structure of an outer convective envelope during their main-sequence. As a result, they will exhibit efficient braking through the effect of magnetised winds and they will most often exhibit long rotation periods of a few days. As a result of their low mass, they exhibit long evolutionary 
timescales of typically a few Gy on the main sequence. This implies 
that, if a given dynamical process is at work in stellar interiors, it 
will have time to leave a durable imprint on the evolution of the star, 
even if its characteristic timescale is relatively long. A good example of such a process is microscopic diffusion, that plays a key role in both solar and stellar modelling. 

Another consequence of the existence of an outer convective envelope in low mass stars is that they will exhibit solar-like oscillations. These oscillations are global acoustic modes of the star that have been measured in both the solar case\cite{Leighton1962}, giving birth to the field of helioseismology, and in the stellar case\cite{Kjeldsen1995,Martic1999,Bouchy2002}, paving the way for the field of asteroseismology of solar-like oscillators. With the advent of space-based photometry missions, the field of asteroseismology experienced an exponential growth, with the detection of global acoustic oscillations in thousands of stars. 

Moreover, the quality of the observations enabled to perform similar analysis to those of the Sun, but on distant stars, increasing the sample of objects for which the internal rotation could be measured. The detection of mixed oscillation modes also gave access to the deep core rotation of evolved stars, providing laboratories for angular momentum transport process at various evolutionary stages. 

In addition, the fact that most solar-like oscillators exhibited slow rotation of their surface allowed modellers to use perturbative approaches applied in helioseismology. 

These measurements proved to be very challenging to explain using classical self-consistent rotating models including the effects of meridional circulation and shear-induced turbulence. While there exists multiple candidates to explain the internal rotation of low-mass stars on the main-sequence, the internal rotation of post-main sequence stars is still impossible to fully reproduce. Behind these difficulties hide the limitations of stellar models, emphasizing the need to improve our physical representation of stellar interiors. 

In this brief review, we will discuss how the internal rotation of low mass stars is measured using a perturbative formalism in the fields of helio- and asteroseismology in Section \ref{Sec:Measure}. We will discuss the implications for angular momentum transport of the seismic constraints obtained on the main-sequence and will discuss the link with the transport of chemicals in Section \ref{Sec:MS}. We will focus here on the constraints from acoustic oscillations only. More complete reviews, discussing also the results of massive gravity mode pulsators can also be found in the litterature\cite{aer19} and classical textbooks also provide a more detailed view and additional references\cite{Maeder2009}. In Section \ref{Sec:PostMS}, we will briefly present the main results using the mixed oscillation modes observed in post-main sequence stars and briefly discuss in Section \ref{Sec:Candi} the implications for the main candidates foreseen as the ``missing angular momentum transport process''. 

\section{Measuring the internal rotation of low mass stars}\label{Sec:Measure}

The surface rotation of stars is accessible in photometry through the observations of spots on the stellar surface, or in spectroscopy through the determination of the rotational broadening of spectral lines. This allows us to test for example the formalisms used for the magnetic braking of low mass stars or the correlation between the depletion of light elements such as lithium or beryllium and the rotation period of stars in clusters. However, to study the internal rotation of stars, the only method available is by using the constraints derived from global oscillation modes. Indeed, their properties are intimately linked to the internal structure and dynamics of stellar interiors. 

In the case of a non-rotating, non-magnetic, isolated star, the global oscillation modes will exhibit spherical symmetry and their angular dependencies can be decomposed on the basis of spherical harmonics. The eigenfunction of an oscillation mode, denoted $\vec{\xi}$, will thus be written in spherical coordinates as
\begin{align}
\vec{\xi}= a(r)Y_{\ell,m}(\theta,\phi) \vec{e}_{r} +b(r)\left[\frac{\partial Y_{\ell,m}(\theta,\phi)}{\partial \theta}\vec{e}_{\theta} + \frac{im Y_{\ell,m}(\theta,\phi)}{\sin \theta}\vec{e}_{\phi}\right], \label{eq:eigenfunction}
\end{align}
with $r$,$\theta$ and $\phi$ the radial and angular coordinates of the system, with their respective unit vectors $\vec{e}_{r}$, $\vec{e}_{\theta}$ and $\vec{e}_{\phi}$, $Y_{\ell,m}$ is the spherical harmonic of degrees $\ell$ and $m$, and $a(r)$ and $b(r)$ are the functions expressing the radial dependencies of the radial and angular components of the eigenfunctions.  
 
There exists actually a degeneracy in the quantum numbers describing the eigenvalues, as they are identified by three quantum numbers, $\ell$, the spherical degree, $n$, the radial order and $m$, the azimuthal degree in the general case, but only two are needed to describe the eigenvalue in the non-rotating case. 

In the case of low-mass stars, we can assume that we are studying a slow rotator. Consequently, the effects of rotation will thus be treated as a perturbation of the operator describing the global acoustic modes, in a similar fashion to the treatment of a weak magnetic field in the case of the hydrogen atom in quantum mechanics. Just as the degeneracy in some eigenstates is lifted by the presence of the magnetic field in the hydrogen atom, the addition of rotation to the eigenvalue problem of stellar oscillations lifts the degeneracy of the eigenvalues, namely the oscillation frequencies of the star. They appear now as multiplets in the oscillation spectrum defined by their three quantum numbers, this time with $m$ taking values between $-\ell$ and $\ell$. This breaking of symmetry stems from the fact that adding rotation to the problem leads to the definition of an equator to the star and that one can now differentiates between the lines of the eigenfunction crossing the equator and those parallel to the equator, that define the quantum numbers $\ell$ and $m$ in the base of spherical harmonics.

In other words, the eigenvalues will now be described as
\begin{align}
\nu_{n,\ell,m}=\nu_{n,\ell,0}+ \delta \nu_{n,\ell,m}
\end{align}

Mathematically, the perturbative approach applied to the stellar oscillation problem leads to a simple integral relation linking the so-called rotational splittings to the internal rotation as function of the radial position $r$ and the latitude $\theta$\cite{Ledoux1951}

\begin{align}
\delta \nu_{n,\ell,m} = m\int_{0}^{R}\int_{0}^{\pi}K_{n,\ell,m}(r,\theta)\Omega(r,\theta)dr d\theta \label{eq:RotSplit}
\end{align}
with $\delta \nu_{n,\ell,m}$ the rotational splitting, $\Omega(r, \theta)$ the internal rotation profile and $K_{n,\ell,m}(r,\theta)$ the so-called kernel function, depending on the internal structure and the eigenfunction of the oscillation mode. 

This integral relation can be solved for the internal rotation profile, using specific numerical techniques (see e.g. \cite{SchouRota}). Given that the linear perturbative approach is valid, the internal rotation profile can be determined independently from the stellar model used to perform the inversion. In other words, a non-rotating stellar model is sufficient to perform an analysis of the internal rotation of the star, if rotation is slow enough to be treated as a perturbation. 

Equation \ref{eq:RotSplit} gives the general $2D$ form of the perturbative relations, but the ability to carry out $2D$ inversions is limited to a few exceptionnal targets, and was only possible for the Sun until very recently. In most seismic analyses, the hypothesis of spherical symmetry will be made and the rotation profile will be inferred as a function of $r$ only. The rotation kernels are then a function of $r$ only and show no explicit dependency in $m$, as shown for example in classical textbooks\cite{Unno1989}. Moreover, the availability of only low $\ell$ modes for asteroseismic targets will also limit the resolution of the inferences. The positions at which the rotational profile can be inferred are intrinsically bound to the physical nature of the observed oscillation modes and the behaviour of their kernel functions. 

\section{Results on the main sequence}\label{Sec:MS}

\begin{figure}
\centering
\includegraphics[width=9cm]{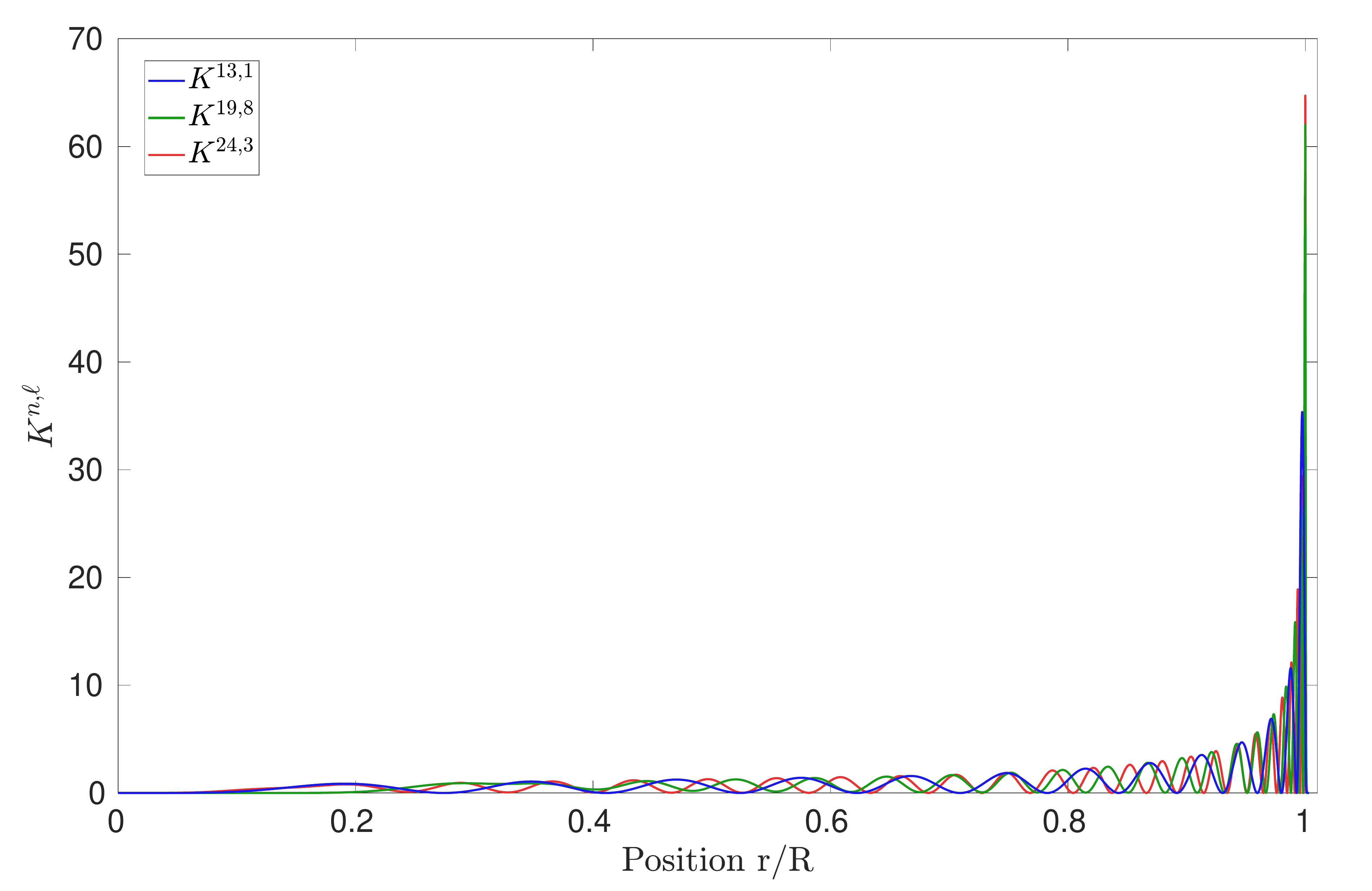}
\caption{$1D$ rotational kernels for various acoustic oscillaton modes of a standard solar model.}
\label{fig:Kernels}
\end{figure}

Main-sequence low mass stars are known to exhibit purely acoustic modes called solar-like oscillations. As indicated by their name, these oscillations have been observed for the first time in the Sun. The main feature of these modes is that they have a much higher amplitude in the outer layers, as indicated in Fig \ref{fig:Kernels} by the $1D$ rotational kernels computed here for a standard solar model. 

Consequently, most of the inferences we will have for all main-sequence solar-like oscillators will focus on the upper radiative layers and the outer convective envelope. The inner $20\%$ of the star will not be accessible to the pressure modes and thus will not be constrained by the inversion procedure. This statement is also true for the Sun, despite the wealth of seismic data accumulated in the last 30 years and has motivated the search for the elusive solar gravity modes. 

\subsection{The solar case}

In the solar case, thousands of oscillation modes with degrees ranging from $\ell=0$ to $\ell=1000$ have been observed \cite{Korzennik2013,Larson2018,Reiter2020}. This tremendous amount of data has allowed a full scan of the internal solar rotation profile, down to $\approx 0.2 R_{\odot}$ (see e.g \cite{Thompson1996}).

The obtained rotation profile exhibits three main features. First, approximately solid-body rotation in the inner radiative layers of the Sun. Second, latitudinal differential rotation in the outer convective zone. Third, these two zones are connected by a narrow region, the so-called tachocline\cite{SpiegelZahn1992,Hughes2007}, where the transition between the two rotation profiles occur. 

Here, we will focus mostly on the properties of the rotation profile in the inner solar radiative zone. The observed nearly uniform rotation of the inner solar layers is actually in complete contradiction with rotating models taking into account the effects of meridional circulation and shear-induced turbulence. Despite attempts at revising the prescriptions for these hydrodynamical processes\cite{mat18}, it appears that their efficiency is too low to prevent the appearence of the characteristic strong differential rotation observed in these models. 

Such a strong differential rotation is in disagreement with observations of the internal rotation of low-mass stars and also leads to a transport of chemicals by the shear-instability in disagreement with helioseismic inversions of the structure of the Sun and the depletion of light elements in low-mass stars. Therefore, additional processes have to been invoked to explain the discrepancies between rotating models and observations. The transport process involved should be very efficient at transporting angular momentum, but not chemical elements. 

The first candidate is a large scale fossil magnetic field \cite{gou98}, the second one are magnetic instabilities \cite{Spruit2002, Eggenberger, Eggenberger2019} and the third one are internal gravity waves\cite{Charbonnel}. We will further discuss the properties of these candidates in Section \ref{Sec:Candi}. They remain, to this day, the main suspects to explain the efficient angular momentum transport in stellar interiors. 

\subsection{Results for \textit{Kepler} and CoRoT targets}

With the advent of the CoRoT and \textit{Kepler} missions, inferences on the internal rotation of main sequence stars could be carried out for a small sample of the best asteroseismic targets. Various authors \cite{lun14,ben15,nie15} have shown that in the case of solar-like oscillators, the measurement of the average rotation seen by acoustic modes was consistent with the surface rotation derived from spots or spectroscopic measurements. In other words, no significant degree of differential rotation was found inside solar-like stars. This again accredited the existence of a very efficient transport of angular momentum acting in these stars, allowing to flatten their internal rotation profile. 

However, a word of caution is required when stating that solar-like stars rotate as "solid-bodies". The constraints derived from asteroseismic data are not able to fully resolve the internal rotation of distant stars, but rather provide an average measurement of their internal rotation. Just as for the solar case, the purely acoustic oscillations observed for main-sequence solar-like oscillators do not allow to probe the deep core of these objects. As we will see in Section \ref{Sec:Candi}, this has important consequences to select one of the three candidates for additional transport mentioned above\footnote{We note that a similar situation is present also in more massive main-sequence stars such as $\gamma$ Doradus stars, where the observed gravity modes only constrain the near core region, but not the convective core itself, as they do not propagate in these regions. Getting constraints on this region requires the analysis of inertial modes, recently observed.}. Recently, two studies\cite{Benomar2018,Bazot2019} have also shown that solar-like oscillators also exhibited latitudinal differential rotation in their convective envelopes. 

\subsection{Link with the transport of chemicals}

An important consequence of the presence of an efficient angular momentum transport mechanism is its potential impact on the transport of chemical elements inside the star. For example, the effects of shear-induced instability in the presence of strong radial differential rotation in the radiative layers will be to erase the effects of microscopic diffusion and inhibit the settling of heavier elements from the convective layers to the deep interior\cite{egg10_sl,dea20}. While this effect should be limited as a result of the very low amplitude of rotation gradient in stellar radiative zones, it might not be fully negligible. In addition, a certain degree of additional mixing is required to reproduce the lithium abundance of the Sun\cite{Richard96Sun} and solar-like stars\cite{Talon1998, egg10_magn, Thevenin17, Lodders2019}. 

\begin{figure}
\centering
\includegraphics[width=7cm]{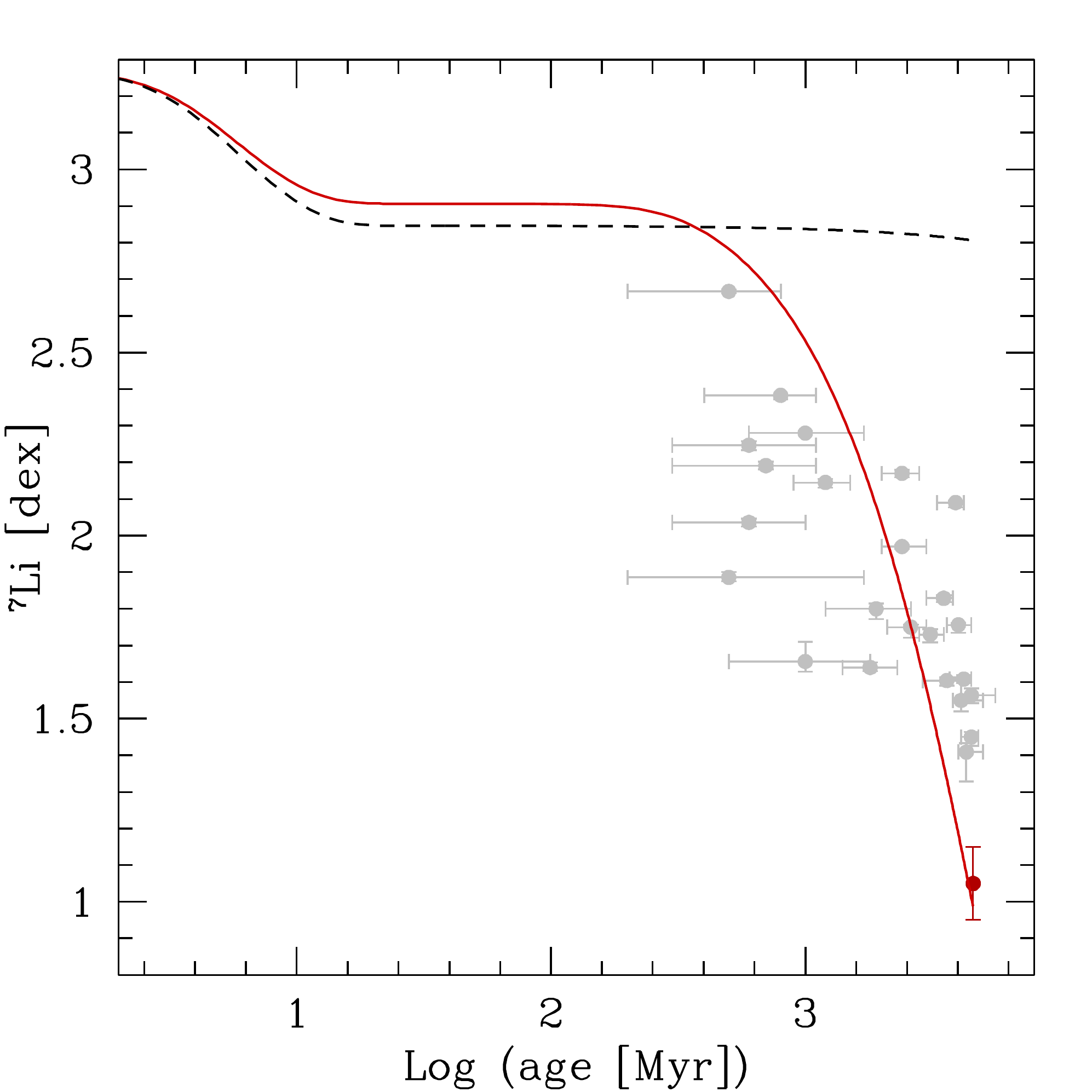}
\caption{Lithium abundance evolution as a function of the logarithm of age. The red dot correspond to the solar Lithium abundance\cite{AGSS09} and the grey dots correspond to the abundance of solar-like stars\cite{car19}.The dashed line indicates the  evolution of the surface Li abundance for a standard solar model, while  the continuous red line corresponds to a model with hydrodynamic and  magnetic instabilities.}
\label{fig:LiEvol}
\end{figure}

This is illustrated in Fig. \ref{fig:LiEvol} for the solar case and young solar-like stars\cite{car19}. The results show that mixing by the shear instability and the effects of magnetic instabilities will lead to a certain degree of mixing at the base of the convective zone thus changing both the helium and lithium abundances. 

Some of the stars observed by the \textit{Kepler} satellite offer excellent testbeds for such effects, as for example the 16Cyg binary system \cite{Deal2015,Buldgen2016,Buldgen2016b,Bazot2020}, where the binarity also allows to test the effects of planetary formation on the abundances of light elements. Moreover, seismic analyses of the helium abundance in the envelope of other \textit{Kepler} targets\cite{Verma2019} have shown the need for a mecanism reducing the efficiency of microscopic diffusion, that could be due to the effects of small rotation gradients.

\section{Results for evolved stars}\label{Sec:PostMS}

The determination of the internal rotation of subgiant and red giant stars\cite{bec12, deh12,deh14,deh15} has been one of the main successes of the space-based photometry missions. These results have been achievable thanks to the observation of so-called mixed oscillation modes. The peculiarity of these modes is that they are of double nature and have high amplitudes both in the deep core and in the outer layers of post-main sequence solar-like oscillators. They exhibit a gravity-like character in the core and an acoustic nature in the envelope. Therefore they can be easily detected and carry information on the deepest layers of evolved stars, regions that are even unaccessible in the Sun due to the purely acoustic nature of the observed oscillations, despite the quality of helioseismic data.

From the analysis of these oscillation modes, it has been shown that subgiant and red giant stars exhibited radial differential rotation \cite{bec12, deh12,deh14,deh15}. Namely, that their cores rotated faster than their outer layers. While this was expected, as a result of the contraction of the core and the expansion of the convective envelope undergone as the star leaves the main sequence, the degree of differential rotation observed is far lower than what is predicted by theoretical models\cite{egg12_rg,mar13,cei13}.

Recent results\cite{Deheuvels2020} have also indicated that young subgiant stars seem to exhibit almost solid-body rotation. Later on in the subgiant phase, they show some degree of differential rotation that develops but it appears that on the red giant phase, the stellar core does not spin up with the evolution\cite{mos12,geh18}. In later stages of evolution such as the red clump or the secondary clump, similar conclusions are drawn regarding the need for a very efficient transport process. 

A crucial point to note here is that to study the efficiency of the missing angular momentum transport process, core rotation is not entirely sufficient, and a measurement of the surface rotation is also required. This is available for a few targets, either from seismology, or from spot modulations. However in the latter case, it seems that the sample is biased towards active, fast rotators, who could perhaps not be quite representative of the bulk of red giants.

The main issue posed by the results obtained for post-main sequence stars is that there is, to this day, no single mechanism capable of explaining all the observations. Analyses of the average efficiency of the missing transport process\cite{egg17, egg19} have shown that it should exhibit a drop in efficiency during the subgiant phase but reincrease as the star ascends the red giant branch. This could possibly point at either a change of regime in the transport process or at multiple transport processes acting at different evolutionary stages. For example, it has been shown that mixed modes themselves would lead to an efficient transport of angular momentum on the upper red giant branch and could potentially take over at stages where another process sees its efficiency reducing\cite{bel15b}. 

\section{Transport mecanism candidates}\label{Sec:Candi}

Apart from the transport by mixed modes, the main suspects under investigation  are the processes already invoked to explain the solar rotation profile. Namely, these are large scale fossil magnetic fields, internal gravity waves and magnetic instabilities.

However, all of them needed to be revised to explain the results observed in post-main sequence stars. We will here briefly summarize the main results and remaining issues with each of these processes. 

\subsection{Fossil magnetic fields}

The first candidate introduced to explain solar rotation is a large-scale fossil field present in the radiative zone of the Sun\cite{gou98}. It was demonstrated that the field, if properly confined, could lead to a rigid rotation profile in the interior of the Sun. 

The main difficulty for the fossil field solution in the solar case is indeed linked to the confinement of the field, as it should not extent in the solar convective zone where differential rotation in latitude is observed. This requires the field to reach a very specific configuration for which simulations give conflictual results regarding its confinement to the radiative zone\cite{bru06,str11,ace13,woo18}. In addition, recent work\cite{spa10} has shown that, in addition to the confinement problem, simulations required a significant increase in the viscosity of the plasma to avoid a dead zone of the field where the profile would not reproduce at all the observed data. 

Nevertheless, a second issue with the fossil field configuration used to explain the solar rotation profile is that it enforces a strictly solid-body rotation in radiative zones. As shown before, this is not what is found in later evolutionary stages and thus another configuration, or another physical process, is required to explain asteroseismic data. One possible configuration for red giants is to have a solid-body rotation profile in radiative zones, combined to a profile in the power law of the radius in convective zones\cite{Kissin2015,Takahashi2021}, of the form $\Omega(r)\propto r^{-\alpha}$, with $\alpha \in \left[1.0, 1.5 \right]$. These studies\cite{Kissin2015,Takahashi2021} showed that such a rotation profile could explain the core rotation rate of red giant branch stars inferred from seismic data\cite{mos12, geh18}.

However, these results are in contradiction with recent studies\cite{dim16,Klion2017,dim18} who seem to indicate that the transition from the slow rotating envelope to the fast rotating core in red giant stars seems to be located in the radiative zone, close to the hydrogen burning shell. A detailed analysis of the case of Kepler 56 was recently performed\cite{Fellay2021}, including a full dedicated modelling of the internal structure of the star and an MCMC analysis of the rotational splittings using simple parametric profiles, including the power law solution provided by large scale fossil fields. Their results show that such a rotation profile cannot be used to explain the individual rotational splittings unless $\alpha$ is allowed to reach higher values of the order of $3.5$, in contradiction with theoretical values\cite{Kissin2015}. This suggests that fossil magnetic fields are not the explanation for the efficient angular momentum transport process acting in evolved stars, and other processes must also be invoked. 

\subsection{Internal gravity waves}

The effect of internal gravity waves has been studied already early on as being of potential interest to the modelling of stellar interiors and applied to the solar case\cite{tal02, Charbonnel}. 

Later studies\cite{Rogers2007,Denissenkov2008} indicated that the process was not efficient enough to actually flatten the solar rotation profile and pointed that the effect of waves excited by Reynold's stresses would leave a strong imprint of the wavefront on the internal solar rotation profile that is incompatible with observations. When applied to later evolutionary stages, the gravity wave model used on the main-sequence\cite{Charbonnel} is also found to be not efficient enough to reproduce the internal rotation of subgiants and red giant stars\cite{ful14}. 

However, hydrodynamical simulations\cite{Rogers2006} underlined the importance of the effects of convective plumes in the generation of gravity waves. These effects were later studied, showing that gravity waves induced by the impact of convective plumes could be much more efficient at transporting angular momentum both in the Sun and subgiant stars\cite{Pincon2016,Pincon2017}. They also showed that it would likely not be able to operate efficiently on the red giant branch, where another process would then be needed. It would then be particularly interesting to compute full evolutionary models with plume-induced gravity waves to investigate their effects on the rotation profile of solar-type and subgiant stars. 

\subsection{Magnetic instabilities}

The last main candidate to explain the efficient transport of angular 
momentum in stellar radiatives zone are magnetic instabilities. Work by H. Spruit\cite{Spruit1999} showed that the first instability likely set in stellar conditions would be the Tayler instability. In the presence of a differentially rotating fluid, the winding up of a seed magnetic field together with the Tayler 
instability has been shown to lead to an efficient angular momentum 
transport mechanism, the so-called Tayler-Spruit dynamo\cite{spr02}. This 
transport mechanism has been shown to be able to reproduce the internal solar rotation profile\cite{Eggenberger, Eggenberger2019}.

While a promising candidate, numerical simulations have shown contrasting results about the apparition of the Tayler-Spruit dynamo\cite{zah07,bra17}. The main difficulty with such conclusions is that such simulations are never in fully realistic stellar conditions. 

Nevertheless, in its original form, the Tayler-Spruit dynamo still proves not efficient enough to counteract the spin up of the core at the end of the main-sequence. Models computed taking it into account still predict faster core rotation rates than the seismically inferred values \cite{can14,den19}. 

\begin{figure}
\centering
\includegraphics[width=9cm]{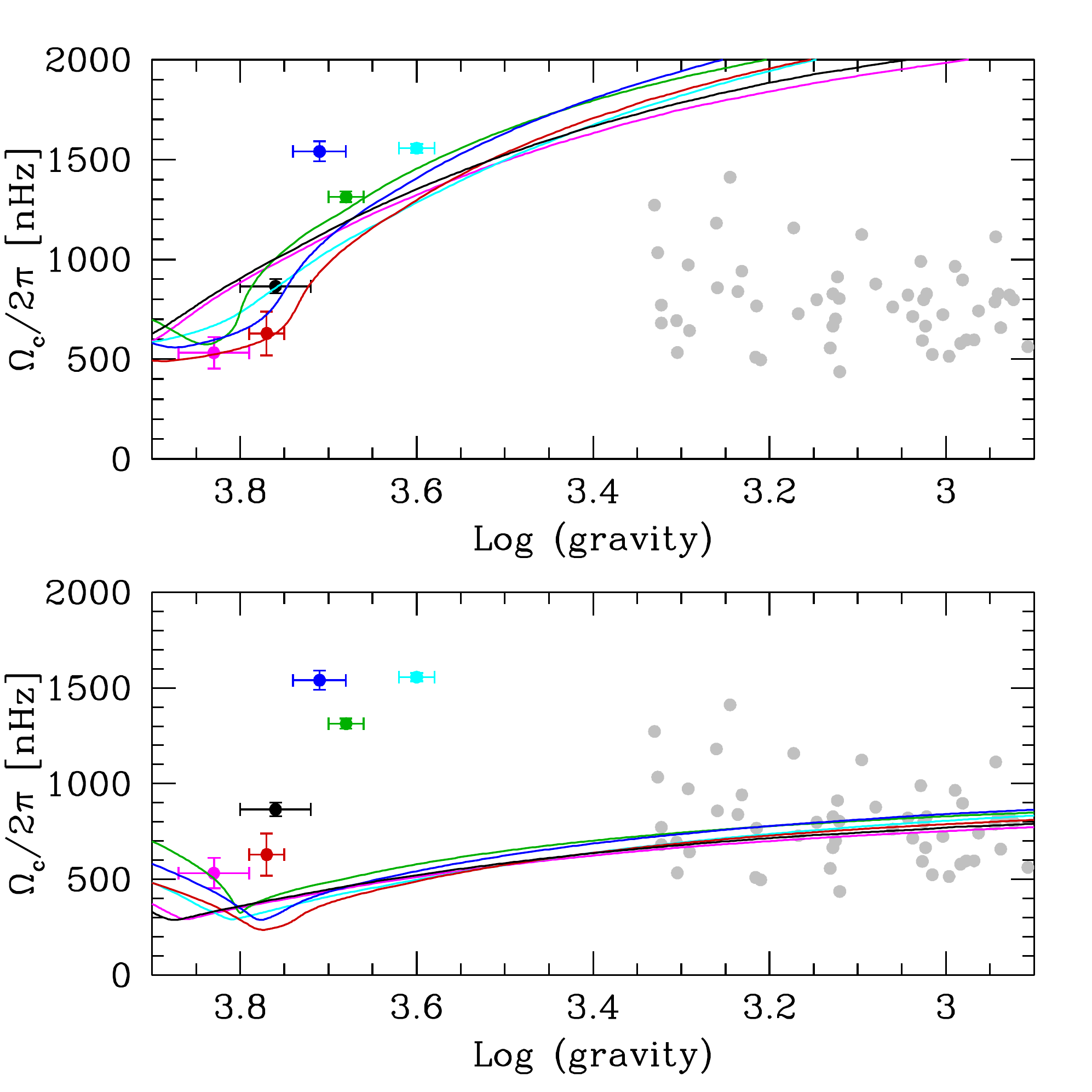}
\caption{Core rotation of subgiant and red giant stars as a function of the logarithm of the surface gravity. The colored points correspond to subgiant stars\cite{deh14} and the grey dots to red giant stars\cite{mos12}. In the upper and lower panels, the calibration parameter $\alpha$ in the revised version of the Tayler-Spruit dynamo is set to $0.5$ and $1.5$, respectively.}
\label{fig:rgcore}
\end{figure}

Recently, a modification of the Tayler-Spruit instability was proposed that led to a more efficient braking of the stellar cores\cite{ful19}. Further analyses\cite{egg19_ful} showed however that this variant could not reproduce all the observational constraints. Namely, it was found to be too efficient for subgiants if one wished to reproduce the red giant cores, or not efficient enough for red giants if the calibration parameter was set to reproduce subgiants, as shown in Fig \ref{fig:rgcore}. This calibration parameter, denoted $\alpha$ has to be varied from $0.5$ to $1.5$ to reproduce respectively the subgiants and the red giants. However, as it is set to the third power in the revised formalism, such a variation is far from anecdotic. A similar issue was found for secondary clump stars and white dwarfs\cite{den20}.

\section{Conclusion}\label{Sec:Conc}

In this brief review, we have described the current state of the vision solar and stellar seismology brought on the internal rotation of low-mass stars in various evolutionary stages. We have focussed on the inferences brought by acoustic and mixed oscillations modes only, trying to provide a complete vision of the current issues at hand regarding the missing angular momentum transport mechanism found to act in stellar radiative zones. 

The formalism for seismic inferences within the first-order perturbative formalism has been introduced in Section \ref{Sec:Measure}, while results for the main-sequence and post-main sequence stars have been discussed in Sections \ref{Sec:MS} and \ref{Sec:PostMS}. We have also briefly introduced the three main candidates currently under extensive investigations in Section \ref{Sec:Candi}. 

The main conclusion to be drawn from the current situation is a sort of stalemate between multiple processes, none of them providing a unifying satisfactory solution. In that respect, further investigations on the effiency of the missing transport process are required, analysing its dependencies with stellar properties such as mass and metallicity. To this end, better characterizing the location of rotation gradients inside stars is crucial, as well as providing both surface and core rotation measurements from a consistent seismic inference technique to fully constrain the efficiency of the transport process. 

It should also be noted that the potential strong influence of chemical composition gradients in the development and efficiency of instabilities (magnetic or hydrodynamic) also calls for a better depiction of the chemical structure of stars. Therefore, the potential solution to the angular momentum transport problem is also tightly linked to the reliability of our vision of the internal structure of stars, that also relies on improving seismic inference techniques. At stakes is likely the recipe for a new generation of solar and stellar models, which would take into account both the effects of rotation, the depletion of light elements and better reproduce seismic and spectroscopic constraints. Moreover, while a unifying solution is tempting, the variations seen at different evolutionary stages might be pointing at various processes taking over during the evolution. To provide a clearer picture, further efforts must be made in the improvement of seismic inference techniques and in the exploitation of the current and future datasets from the CoRoT \cite{bag06}, \textit{Kepler} \cite{bor10}, TESS \cite{Ricker2015} and Plato \cite{Rauer2014} missions.

\bibliographystyle{ws-procs961x669}
\bibliography{BiblioProcWG16}

\begin{thebibliography}{10}

\bibitem{Leighton1962}
R.~B. {Leighton}, R.~W. {Noyes} and G.~W. {Simon}, {Velocity Fields in the
  Solar Atmosphere. I. Preliminary Report.}, {\em ApJ} {\bf 135}, p. 474 (March
  1962).

\bibitem{Kjeldsen1995}
H.~{Kjeldsen}, T.~R. {Bedding}, M.~{Viskum} and S.~{Frandsen}, {Solarlike
  Oscillations in eta Boo}, {\em AJ} {\bf 109}, p. 1313 (March 1995).

\bibitem{Martic1999}
M.~{Marti{\'c}}, J.~{Schmitt}, J.~C. {Lebrun}, C.~{Barban}, P.~{Connes},
  F.~{Bouchy}, E.~{Michel}, A.~{Baglin}, T.~{Appourchaux} and J.~L. {Bertaux},
  {Evidence for global pressure oscillations on Procyon}, {\em A\&A} {\bf 351},
  993 (November 1999).

\bibitem{Bouchy2002}
F.~{Bouchy} and F.~{Carrier}, {The acoustic spectrum of alpha Cen A}, {\em
  A\&A} {\bf 390}, 205 (July 2002).

\bibitem{aer19}
C.~{Aerts}, S.~{Mathis} and T.~M. {Rogers}, {Angular Momentum Transport in
  Stellar Interiors}, {\em Annu. Rev. Astron. Astrophys.} {\bf 57}, 35 (August
  2019).

\bibitem{Maeder2009}
A.~{Maeder}, {\em {Physics, Formation and Evolution of Rotating Stars}} 2009.

\bibitem{Ledoux1951}
P.~{Ledoux}, {The Nonradial Oscillations of Gaseous Stars and the Problem of
  Beta Canis Majoris.}, {\em ApJ} {\bf 114}, p. 373 (November 1951).

\bibitem{SchouRota}
J.~{Schou}, H.~M. {Antia}, S.~{Basu}, R.~S. {Bogart}, R.~I. {Bush}, S.~M.
  {Chitre}, J.~{Christensen-Dalsgaard}, M.~P. {Di Mauro}, W.~A. {Dziembowski},
  A.~{Eff-Darwich}, D.~O. {Gough}, D.~A. {Haber}, J.~T. {Hoeksema}, R.~{Howe},
  S.~G. {Korzennik}, A.~G. {Kosovichev}, R.~M. {Larsen}, F.~P. {Pijpers}, P.~H.
  {Scherrer}, T.~{Sekii}, T.~D. {Tarbell}, A.~M. {Title}, M.~J. {Thompson} and
  J.~{Toomre}, {Helioseismic Studies of Differential Rotation in the Solar
  Envelope by the Solar Oscillations Investigation Using the Michelson Doppler
  Imager}, {\em ApJ} {\bf 505}, 390 (September 1998).

\bibitem{Unno1989}
W.~{Unno}, Y.~{Osaki}, H.~{Ando}, H.~{Saio} and H.~{Shibahashi}, {\em
  {Nonradial oscillations of stars}} 1989.

\bibitem{Korzennik2013}
S.~G. {Korzennik}, M.~C. {Rabello-Soares}, J.~{Schou} and T.~P. {Larson},
  {Accurate Characterization of High-degree Modes Using MDI Observations}, {\em
  ApJ} {\bf 772}, p.~87 (August 2013).

\bibitem{Larson2018}
T.~P. {Larson} and J.~{Schou}, {Global-Mode Analysis of Full-Disk Data from the
  Michelson Doppler Imager and the Helioseismic and Magnetic Imager}, {\em
  Solar Physics} {\bf 293}, p.~29 (February 2018).

\bibitem{Reiter2020}
J.~{Reiter}, J.~{Rhodes}, E.~J., A.~G. {Kosovichev}, P.~H. {Scherrer}, T.~P.
  {Larson} and I.~{}, S.~F.~Pinkerton, {A Method for the Estimation of f- and
  p-mode Parameters and Rotational Splitting Coefficients from Un-averaged
  Solar Oscillation Power Spectra}, {\em ApJ} {\bf 894}, p.~80 (May 2020).

\bibitem{Thompson1996}
M.~J. {Thompson}, J.~{Toomre}, E.~R. {Anderson}, H.~M. {Antia},
  G.~{Berthomieu}, D.~{Burtonclay}, S.~M. {Chitre}, J.~{Christensen-Dalsgaard},
  T.~{Corbard}, M.~{De Rosa}, C.~R. {Genovese}, D.~O. {Gough}, D.~A. {Haber},
  J.~W. {Harvey}, F.~{Hill}, R.~{Howe}, S.~G. {Korzennik}, A.~G. {Kosovichev},
  J.~W. {Leibacher}, F.~P. {Pijpers}, J.~{Provost}, J.~{Rhodes}, E.~J.,
  J.~{Schou}, T.~{Sekii}, P.~B. {Stark} and P.~R. {Wilson}, {Differential
  Rotation and Dynamics of the Solar Interior}, {\em Science} {\bf 272}, 1300
  (May 1996).

\bibitem{SpiegelZahn1992}
E.~A. {Spiegel} and J.-P. {Zahn}, {The solar tachocline}, {\em A\&Ap} {\bf
  265}, 106 (November 1992).

\bibitem{Hughes2007}
D.~W. {Hughes}, R.~{Rosner} and N.~O. {Weiss}, {\em {The Solar Tachocline}} May
  2007.

\bibitem{mat18}
S.~{Mathis}, V.~{Prat}, L.~{Amard}, C.~{Charbonnel}, A.~{Palacios},
  N.~{Lagarde} and P.~{Eggenberger}, {Anisotropic turbulent transport in stably
  stratified rotating stellar radiation zones}, {\em A\&A} {\bf 620}, p. A22
  (December 2018).

\bibitem{gou98}
D.~O. {Gough} and M.~E. {McIntyre}, {Inevitability of a magnetic field in the
  Sun's radiative interior}, {\em Nature} {\bf 394}, 755 (August 1998).

\bibitem{Spruit2002}
H.~C. {Spruit}, {Dynamo action by differential rotation in a stably stratified
  stellar interior}, {\em A\&A} {\bf 381}, 923 (January 2002).

\bibitem{Eggenberger}
P.~{Eggenberger}, A.~{Maeder} and G.~{Meynet}, {Stellar evolution with rotation
  and magnetic fields. IV. The solar rotation profile}, {\em A\&A} {\bf 440},
  L9 (September 2005).

\bibitem{Eggenberger2019}
P.~{Eggenberger}, G.~{Buldgen} and S.~J.~A.~J. {Salmon}, {Rotation rate of the
  solar core as a key constraint to magnetic angular momentum transport in
  stellar interiors}, {\em A\&A} {\bf 626}, p.~L1 (Jun 2019).

\bibitem{Charbonnel}
C.~{Charbonnel} and S.~{Talon}, {Influence of Gravity Waves on the Internal
  Rotation and Li Abundance of Solar-Type Stars}, {\em Science} {\bf 309}, 2189
  (September 2005).

\bibitem{lun14}
M.~N. {Lund}, M.~S. {Miesch} and J.~{Christensen-Dalsgaard}, {Differential
  Rotation in Main-sequence Solar-like Stars: Qualitative Inference from
  Asteroseismic Data}, {\em ApJ} {\bf 790}, p. 121 (August 2014).

\bibitem{ben15}
O.~{Benomar}, M.~{Takata}, H.~{Shibahashi}, T.~{Ceillier} and R.~A.
  {Garc{\'{\i}}a}, {Nearly uniform internal rotation of solar-like
  main-sequence stars revealed by space-based asteroseismology and
  spectroscopic measurements}, {\em MNRAS} {\bf 452}, 2654 (September 2015).

\bibitem{nie15}
M.~B. {Nielsen}, H.~{Schunker}, L.~{Gizon} and W.~H. {Ball}, {Constraining
  differential rotation of Sun-like stars from asteroseismic and starspot
  rotation periods}, {\em A\&A} {\bf 582}, p. A10 (October 2015).

\bibitem{Benomar2018}
O.~{Benomar}, M.~{Bazot}, M.~B. {Nielsen}, L.~{Gizon}, T.~{Sekii}, M.~{Takata},
  H.~{Hotta}, S.~{Hanasoge}, K.~R. {Sreenivasan} and
  J.~{Christensen-Dalsgaard}, {Asteroseismic detection of latitudinal
  differential rotation in 13 Sun-like stars}, {\em Science} {\bf 361}, 1231
  (September 2018).

\bibitem{Bazot2019}
M.~{Bazot}, O.~{Benomar}, J.~{Christensen-Dalsgaard}, L.~{Gizon},
  S.~{Hanasoge}, M.~{Nielsen}, P.~{Petit} and K.~R. {Sreenivasan}, {Latitudinal
  differential rotation in the solar analogues 16 Cygni A and B}, {\em A\&A}
  {\bf 623}, p. A125 (March 2019).

\bibitem{egg10_sl}
P.~{Eggenberger}, G.~{Meynet}, A.~{Maeder}, A.~{Miglio}, J.~{Montalban},
  F.~{Carrier}, S.~{Mathis}, C.~{Charbonnel} and S.~{Talon}, {Effects of
  rotational mixing on the asteroseismic properties of solar-type stars}, {\em
  A\&A} {\bf 519}, p. A116 (September 2010).

\bibitem{dea20}
M.~{Deal}, M.~J. {Goupil}, J.~P. {Marques}, D.~R. {Reese} and Y.~{Lebreton},
  {Chemical mixing in low mass stars. I. Rotation against atomic diffusion
  including radiative acceleration}, {\em A\&A} {\bf 633}, p. A23 (January
  2020).

\bibitem{Richard96Sun}
O.~{Richard}, S.~{Vauclair}, C.~{Charbonnel} and W.~A. {Dziembowski}, {New
  solar models including helioseismological constraints and light-element
  depletion.}, {\em A\&Ap} {\bf 312}, 1000 (August 1996).

\bibitem{Talon1998}


\bibitem{egg10_magn}
P.~{Eggenberger}, A.~{Maeder} and G.~{Meynet}, {Effects of rotation and
  magnetic fields on the lithium abundance and asteroseismic properties of
  exoplanet-host stars}, {\em A\&A} {\bf 519}, p.~L2 (September 2010).

\bibitem{Thevenin17}
F.~{Th{\'e}venin}, A.~V. {Oreshina}, V.~A. {Baturin}, A.~B. {Gorshkov},
  P.~{Morel} and J.~{Provost}, {Evolution of lithium abundance in the Sun and
  solar twins}, {\em A\&A} {\bf 598}, p. A64 (February 2017).

\bibitem{Lodders2019}
M.~{Carlos}, J.~{Mel{\'e}ndez}, L.~{Spina}, L.~A. {dos Santos}, M.~{Bedell},
  I.~{Ramirez}, M.~{Asplund}, J.~L. {Bean}, D.~{Yong}, J.~{Yana Galarza} and
  A.~{Alves-Brito}, {The Li-age correlation: the Sun is unusually Li deficient
  for its age}, {\em MNRAS} {\bf 485}, 4052 (May 2019).

\bibitem{car19}
M.~{Carlos}, J.~{Mel{\'e}ndez}, L.~{Spina}, L.~A. {dos Santos}, M.~{Bedell},
  I.~{Ramirez}, M.~{Asplund}, J.~L. {Bean}, D.~{Yong}, J.~{Yana Galarza} and
  A.~{Alves-Brito}, {The Li-age correlation: the Sun is unusually Li deficient
  for its age}, {\em MNRAS} {\bf 485}, 4052 (May 2019).

\bibitem{AGSS09}
M.~{Asplund}, N.~{Grevesse}, A.~J. {Sauval} and P.~{Scott}, {The Chemical
  Composition of the Sun}, {\em ARA\&A} {\bf 47}, 481 (September 2009).

\bibitem{Deal2015}
M.~{Deal}, O.~{Richard} and S.~{Vauclair}, {Accretion of planetary matter and
  the lithium problem in the 16 Cygni stellar system}, {\em A\&A} {\bf 584}, p.
  A105 (December 2015).

\bibitem{Buldgen2016}
G.~{Buldgen}, D.~R. {Reese} and M.~A. {Dupret}, {Constraints on the structure
  of 16 Cygni A and 16 Cygni B using inversion techniques}, {\em A\&A} {\bf
  585}, p. A109 (January 2016).

\bibitem{Buldgen2016b}
G.~{Buldgen}, S.~J.~A.~J. {Salmon}, D.~R. {Reese} and M.~A. {Dupret}, {In-depth
  study of 16CygB using inversion techniques}, {\em A\&A} {\bf 596}, p. A73
  (December 2016).

\bibitem{Bazot2020}
M.~{Bazot}, {Uncertainties and biases in modelling 16 Cygni A and B}, {\em
  A\&A} {\bf 635}, p. A26 (March 2020).

\bibitem{Verma2019}
K.~{Verma} and V.~{Silva Aguirre}, {Helium settling in F stars: constraining
  turbulent mixing using observed helium glitch signature}, {\em MNRAS} {\bf
  489}, 1850 (October 2019).

\bibitem{bec12}
P.~G. {Beck}, J.~{Montalban}, T.~{Kallinger}, J.~{De Ridder}, C.~{Aerts}, R.~A.
  {Garc{\'{\i}}a}, S.~{Hekker}, M.-A. {Dupret}, B.~{Mosser}, P.~{Eggenberger},
  D.~{Stello}, Y.~{Elsworth}, S.~{Frandsen}, F.~{Carrier}, M.~{Hillen},
  M.~{Gruberbauer}, J.~{Christensen-Dalsgaard}, A.~{Miglio}, M.~{Valentini},
  T.~R. {Bedding}, H.~{Kjeldsen}, F.~R. {Girouard}, J.~R. {Hall} and K.~A.
  {Ibrahim}, {Fast core rotation in red-giant stars as revealed by
  gravity-dominated mixed modes}, {\em Nature} {\bf 481}, 55 (January 2012).

\bibitem{deh12}
S.~{Deheuvels}, R.~A. {Garc{\'{\i}}a}, W.~J. {Chaplin}, S.~{Basu}, H.~M.
  {Antia}, T.~{Appourchaux}, O.~{Benomar}, G.~R. {Davies}, Y.~{Elsworth},
  L.~{Gizon}, M.~J. {Goupil}, D.~R. {Reese}, C.~{Regulo}, J.~{Schou},
  T.~{Stahn}, L.~{Casagrande}, J.~{Christensen-Dalsgaard}, D.~{Fischer},
  S.~{Hekker}, H.~{Kjeldsen}, S.~{Mathur}, B.~{Mosser}, M.~{Pinsonneault},
  J.~{Valenti}, J.~L. {Christiansen}, K.~{Kinemuchi} and F.~{Mullally},
  {Seismic Evidence for a Rapidly Rotating Core in a Lower-giant-branch Star
  Observed with Kepler}, {\em ApJ} {\bf 756}, p.~19 (September 2012).

\bibitem{deh14}
S.~{Deheuvels}, G.~{Do{\u g}an}, M.~J. {Goupil}, T.~{Appourchaux},
  O.~{Benomar}, H.~{Bruntt}, T.~L. {Campante}, L.~{Casagrande}, T.~{Ceillier},
  G.~R. {Davies}, P.~{De Cat}, J.~N. {Fu}, R.~A. {Garc{\'{\i}}a}, A.~{Lobel},
  B.~{Mosser}, D.~R. {Reese}, C.~{Regulo}, J.~{Schou}, T.~{Stahn}, A.~O.
  {Thygesen}, X.~H. {Yang}, W.~J. {Chaplin}, J.~{Christensen-Dalsgaard},
  P.~{Eggenberger}, L.~{Gizon}, S.~{Mathis}, J.~{Molenda-{\.Z}akowicz} and
  M.~{Pinsonneault}, {Seismic constraints on the radial dependence of the
  internal rotation profiles of six Kepler subgiants and young red giants},
  {\em A\&A} {\bf 564}, p. A27 (April 2014).

\bibitem{deh15}
S.~{Deheuvels}, J.~{Ballot}, P.~G. {Beck}, B.~{Mosser}, R.~{{\O}stensen}, R.~A.
  {Garc{\'{\i}}a} and M.~J. {Goupil}, {Seismic evidence for a weak radial
  differential rotation in intermediate-mass core helium burning stars}, {\em
  A\&A} {\bf 580}, p. A96 (August 2015).

\bibitem{egg12_rg}
P.~{Eggenberger}, J.~{Montalb{\'a}n} and A.~{Miglio}, {Angular momentum
  transport in stellar interiors constrained by rotational splittings of mixed
  modes in red giants}, {\em A\&A} {\bf 544}, p.~L4 (August 2012).

\bibitem{mar13}
J.~P. {Marques}, M.~J. {Goupil}, Y.~{Lebreton}, S.~{Talon}, A.~{Palacios},
  K.~{Belkacem}, R.-M. {Ouazzani}, B.~{Mosser}, A.~{Moya}, P.~{Morel},
  B.~{Pichon}, S.~{Mathis}, J.-P. {Zahn}, S.~{Turck-Chi{\`e}ze} and P.~A.~P.
  {Nghiem}, {Seismic diagnostics for transport of angular momentum in stars. I.
  Rotational splittings from the pre-main sequence to the red-giant branch},
  {\em A\&A} {\bf 549}, p. A74 (January 2013).

\bibitem{cei13}
T.~{Ceillier}, P.~{Eggenberger}, R.~A. {Garc{\'{\i}}a} and S.~{Mathis},
  {Understanding angular momentum transport in red giants: the case of KIC
  7341231}, {\em A\&A} {\bf 555}, p. A54 (July 2013).

\bibitem{Deheuvels2020}
S.~{Deheuvels}, J.~{Ballot}, P.~{Eggenberger}, F.~{Spada}, A.~{Noll} and J.~W.
  {den Hartogh}, {Seismic evidence for near solid-body rotation in two Kepler
  subgiants and implications for angular momentum transport}, {\em A\&A} {\bf
  641}, p. A117 (September 2020).

\bibitem{mos12}
B.~{Mosser}, M.~J. {Goupil}, K.~{Belkacem}, J.~P. {Marques}, P.~G. {Beck},
  S.~{Bloemen}, J.~{De Ridder}, C.~{Barban}, S.~{Deheuvels}, Y.~{Elsworth},
  S.~{Hekker}, T.~{Kallinger}, R.~M. {Ouazzani}, M.~{Pinsonneault},
  R.~{Samadi}, D.~{Stello}, R.~A. {Garc{\'{\i}}a}, T.~C. {Klaus}, J.~{Li},
  S.~{Mathur} and R.~L. {Morris}, {Spin down of the core rotation in red
  giants}, {\em A\&A} {\bf 548}, p. A10 (December 2012).

\bibitem{geh18}
C.~{Gehan}, B.~{Mosser}, E.~{Michel}, R.~{Samadi} and T.~{Kallinger}, {Core
  rotation braking on the red giant branch for various mass ranges}, {\em A\&A}
  {\bf 616}, p. A24 (Aug 2018).

\bibitem{egg17}
P.~{Eggenberger}, N.~{Lagarde}, A.~{Miglio}, J.~{Montalb{\'a}n},
  S.~{Ekstr{\"o}m}, C.~{Georgy}, G.~{Meynet}, S.~{Salmon}, T.~{Ceillier}, R.~A.
  {Garc{\'{\i}}a}, S.~{Mathis}, S.~{Deheuvels}, A.~{Maeder}, J.~W. {den
  Hartogh} and R.~{Hirschi}, {Constraining the efficiency of angular momentum
  transport with asteroseismology of red giants: the effect of stellar mass},
  {\em A\&A} {\bf 599}, p. A18 (March 2017).

\bibitem{egg19}
P.~{Eggenberger}, S.~{Deheuvels}, A.~{Miglio}, S.~{Ekstr{\"o}m}, C.~{Georgy},
  G.~{Meynet}, N.~{Lagarde}, S.~{Salmon}, G.~{Buldgen}, J.~{Montalb{\'a}n},
  F.~{Spada} and J.~{Ballot}, {Asteroseismology of evolved stars to constrain
  the internal transport of angular momentum. I. Efficiency of transport during
  the subgiant phase}, {\em A\&A} {\bf 621}, p. A66 (January 2019).

\bibitem{bel15b}
K.~{Belkacem}, J.~P. {Marques}, M.~J. {Goupil}, B.~{Mosser}, T.~{Sonoi}, R.~M.
  {Ouazzani}, M.~A. {Dupret}, S.~{Mathis} and M.~{Grosjean}, {Angular momentum
  redistribution by mixed modes in evolved low-mass stars. II. Spin-down of the
  core of red giants induced by mixed modes}, {\em A\&A} {\bf 579}, p. A31
  (July 2015).

\bibitem{bru06}
A.~S. {Brun} and J.-P. {Zahn}, {Magnetic confinement of the solar tachocline},
  {\em A\&A} {\bf 457}, 665 (October 2006).

\bibitem{str11}
A.~{Strugarek}, A.~S. {Brun} and J.-P. {Zahn}, {Magnetic confinement of the
  solar tachocline: II. Coupling to a convection zone}, {\em A\&A} {\bf 532},
  p. A34 (August 2011).

\bibitem{ace13}
L.~A. {Acevedo-Arreguin}, P.~{Garaud} and T.~S. {Wood}, {Dynamics of the solar
  tachocline - III. Numerical solutions of the Gough and McIntyre model}, {\em
  MNRAS} {\bf 434}, 720 (September 2013).

\bibitem{woo18}
T.~S. {Wood} and N.~H. {Brummell}, {A Self-consistent Model of the Solar
  Tachocline}, {\em ApJ} {\bf 853}, p.~97 (February 2018).

\bibitem{spa10}
F.~{Spada}, A.~C. {Lanzafame} and A.~F. {Lanza}, {A semi-analytic approach to
  angular momentum transport in stellar radiative interiors}, {\em MNRAS} {\bf
  404}, 641 (May 2010).

\bibitem{Kissin2015}
Y.~{Kissin} and C.~{Thompson}, {Rotation of Giant Stars}, {\em ApJ} {\bf 808},
  p.~35 (July 2015).

\bibitem{Takahashi2021}
K.~{Takahashi} and N.~{Langer}, {Modeling of magneto-rotational stellar
  evolution. I. Method and first applications}, {\em A\&A} {\bf 646}, p. A19
  (February 2021).

\bibitem{dim16}
M.~P. {Di Mauro}, R.~{Ventura}, D.~{Cardini}, D.~{Stello},
  J.~{Christensen-Dalsgaard}, W.~A. {Dziembowski}, L.~{Patern{\`o}}, P.~G.
  {Beck}, S.~{Bloemen}, G.~R. {Davies}, K.~{De Smedt}, Y.~{Elsworth}, R.~A.
  {Garc{\'{\i}}a}, S.~{Hekker}, B.~{Mosser} and A.~{Tkachenko}, {Internal
  Rotation of the Red-giant Star KIC 4448777 by Means of Asteroseismic
  Inversion}, {\em ApJ} {\bf 817}, p.~65 (January 2016).

\bibitem{Klion2017}
H.~{Klion} and E.~{Quataert}, {A diagnostic for localizing red giant
  differential rotation}, {\em MNRAS} {\bf 464}, L16 (January 2017).

\bibitem{dim18}
M.~P. {Di Mauro}, R.~{Ventura}, E.~{Corsaro} and B.~{Lustosa De Moura}, {The
  Rotational Shear Layer inside the Early Red-giant Star KIC 4448777}, {\em
  ApJ} {\bf 862}, p.~9 (July 2018).

\bibitem{Fellay2021}
L.~{Fellay}, G.~{Buldgen}, P.~{Eggenberger}, S.~{Khan}, S.~J.~A.~J. {Salmon},
  A.~{Miglio} and J.~{Montalb{\'a}n}, {Asteroseismology of evolved stars to
  constrain the internal transport of angular momentum. IV. Internal rotation
  of Kepler 56 from an MCMC analysis of the rotational splittings}, {\em arXiv
  e-prints} , p. arXiv:2108.02670 (August 2021).

\bibitem{tal02}
S.~{Talon}, P.~{Kumar} and J.-P. {Zahn}, {Angular Momentum Extraction by
  Gravity Waves in the Sun}, {\em ApJL} {\bf 574}, L175 (August 2002).

\bibitem{Rogers2007}
T.~M. {Rogers}, {Numerical Simulations of Gravity Wave Driven Shear Flows in
  the Solar Tachocline}, in {\em Unsolved Problems in Stellar Physics: A
  Conference in Honor of Douglas Gough\/},  eds. R.~J. {Stancliffe},
  G.~{Houdek}, R.~G. {Martin} and C.~A. {Tout}, American Institute of Physics
  Conference Series, Vol.~948November 2007.

\bibitem{Denissenkov2008}
P.~A. {Denissenkov}, M.~{Pinsonneault} and K.~B. {MacGregor}, {What Prevents
  Internal Gravity Waves from Disturbing the Solar Uniform Rotation?}, {\em
  ApJ} {\bf 684}, 757 (September 2008).

\bibitem{ful14}
J.~{Fuller}, D.~{Lecoanet}, M.~{Cantiello} and B.~{Brown}, {Angular Momentum
  Transport via Internal Gravity Waves in Evolving Stars}, {\em ApJ} {\bf 796},
  p.~17 (November 2014).

\bibitem{Rogers2006}
T.~M. {Rogers} and G.~A. {Glatzmaier}, {Angular Momentum Transport by Gravity
  Waves in the Solar Interior}, {\em ApJ} {\bf 653}, 756 (December 2006).

\bibitem{Pincon2016}
C.~{Pin{\c{c}}on}, K.~{Belkacem} and M.~J. {Goupil}, {Generation of internal
  gravity waves by penetrative convection}, {\em A\&A} {\bf 588}, p. A122
  (April 2016).

\bibitem{Pincon2017}
C.~{Pin{\c{c}}on}, K.~{Belkacem}, M.~J. {Goupil} and J.~P. {Marques}, {Can
  plume-induced internal gravity waves regulate the core rotation of subgiant
  stars?}, {\em A\&A} {\bf 605}, p. A31 (September 2017).

\bibitem{Spruit1999}
H.~C. {Spruit}, {Differential rotation and magnetic fields in stellar
  interiors}, {\em A\&A} {\bf 349}, 189 (September 1999).

\bibitem{spr02}
H.~C. {Spruit}, {Dynamo action by differential rotation in a stably stratified
  stellar interior}, {\em A\&A} {\bf 381}, 923 (January 2002).

\bibitem{zah07}
J.~{Zahn}, A.~S. {Brun} and S.~{Mathis}, {On magnetic instabilities and dynamo
  action in stellar radiation zones}, {\em A\&A} {\bf 474}, 145 (October 2007).

\bibitem{bra17}
J.~{Braithwaite} and H.~C. {Spruit}, {Magnetic fields in non-convective regions
  of stars}, {\em Royal Society Open Science} {\bf 4}, p. 160271 (February
  2017).

\bibitem{can14}
M.~{Cantiello}, C.~{Mankovich}, L.~{Bildsten}, J.~{Christensen-Dalsgaard} and
  B.~{Paxton}, {Angular Momentum Transport within Evolved Low-mass Stars}, {\em
  ApJ} {\bf 788}, p.~93 (June 2014).

\bibitem{den19}
J.~W. {den Hartogh}, P.~{Eggenberger} and R.~{Hirschi}, {Constraining transport
  of angular momentum in stars. Combining asteroseismic observations of core
  helium burning stars and white dwarfs}, {\em A\&A} {\bf 622}, p. A187
  (February 2019).

\bibitem{ful19}
J.~{Fuller}, A.~L. {Piro} and A.~S. {Jermyn}, {Slowing the spins of stellar
  cores}, {\em MNRAS} {\bf 485}, 3661 (May 2019).

\bibitem{egg19_ful}
P.~{Eggenberger}, J.~W. {den Hartogh}, G.~{Buldgen}, G.~{Meynet}, S.~J.~A.~J.
  {Salmon} and S.~{Deheuvels}, {Asteroseismology of evolved stars to constrain
  the internal transport of angular momentum. II. Test of a revised
  prescription for transport by the Tayler instability}, {\em A\&A} {\bf 631},
  p.~L6 (November 2019).

\bibitem{den20}
J.~W. {den Hartogh}, P.~{Eggenberger} and S.~{Deheuvels}, {Asteroseismology of
  evolved stars to constrain the internal transport of angular momentum. III.
  Using the rotation rates of intermediate-mass stars to test the
  Fuller-formalism}, {\em A\&A} {\bf 634}, p. L16 (February 2020).

\bibitem{bag06}
A.~{Baglin}, M.~{Auvergne}, L.~{Boisnard}, T.~{Lam-Trong}, P.~{Barge},
  C.~{Catala}, M.~{Deleuil}, E.~{Michel} and W.~{Weiss}, {CoRoT: a high
  precision photometer for stellar ecolution and exoplanet finding}, in {\em
  36th COSPAR Scientific Assembly\/}, , COSPAR Meeting Vol.~362006.

\bibitem{bor10}
W.~J. {Borucki}, D.~{Koch}, G.~{Basri}, N.~{Batalha}, T.~{Brown},
  D.~{Caldwell}, J.~{Caldwell}, J.~{Christensen-Dalsgaard}, W.~D. {Cochran},
  E.~{DeVore}, E.~W. {Dunham}, A.~K. {Dupree}, T.~N. {Gautier}, J.~C. {Geary},
  R.~{Gilliland}, A.~{Gould}, S.~B. {Howell}, J.~M. {Jenkins}, Y.~{Kondo},
  D.~W. {Latham}, G.~W. {Marcy}, S.~{Meibom}, H.~{Kjeldsen}, J.~J. {Lissauer},
  D.~G. {Monet}, D.~{Morrison}, D.~{Sasselov}, J.~{Tarter}, A.~{Boss},
  D.~{Brownlee}, T.~{Owen}, D.~{Buzasi}, D.~{Charbonneau}, L.~{Doyle},
  J.~{Fortney}, E.~B. {Ford}, M.~J. {Holman}, S.~{Seager}, J.~H. {Steffen},
  W.~F. {Welsh}, J.~{Rowe}, H.~{Anderson}, L.~{Buchhave}, D.~{Ciardi},
  L.~{Walkowicz}, W.~{Sherry}, E.~{Horch}, H.~{Isaacson}, M.~E. {Everett},
  D.~{Fischer}, G.~{Torres}, J.~A. {Johnson}, M.~{Endl}, P.~{MacQueen}, S.~T.
  {Bryson}, J.~{Dotson}, M.~{Haas}, J.~{Kolodziejczak}, J.~{Van Cleve},
  H.~{Chandrasekaran}, J.~D. {Twicken}, E.~V. {Quintana}, B.~D. {Clarke},
  C.~{Allen}, J.~{Li}, H.~{Wu}, P.~{Tenenbaum}, E.~{Verner}, F.~{Bruhweiler},
  J.~{Barnes} and A.~{Prsa}, {Kepler Planet-Detection Mission: Introduction and
  First Results}, {\em Science} {\bf 327}, 977 (February 2010).

\bibitem{Ricker2015}
G.~R. {Ricker}, J.~N. {Winn}, R.~{Vanderspek}, D.~W. {Latham}, G.~{\'A}.
  {Bakos}, J.~L. {Bean}, Z.~K. {Berta-Thompson}, T.~M. {Brown}, L.~{Buchhave},
  N.~R. {Butler}, R.~P. {Butler}, W.~J. {Chaplin}, D.~{Charbonneau},
  J.~{Christensen-Dalsgaard}, M.~{Clampin}, D.~{Deming}, J.~{Doty}, N.~{De
  Lee}, C.~{Dressing}, E.~W. {Dunham}, M.~{Endl}, F.~{Fressin}, J.~{Ge},
  T.~{Henning}, M.~J. {Holman}, A.~W. {Howard}, S.~{Ida}, J.~M. {Jenkins},
  G.~{Jernigan}, J.~A. {Johnson}, L.~{Kaltenegger}, N.~{Kawai}, H.~{Kjeldsen},
  G.~{Laughlin}, A.~M. {Levine}, D.~{Lin}, J.~J. {Lissauer}, P.~{MacQueen},
  G.~{Marcy}, P.~R. {McCullough}, T.~D. {Morton}, N.~{Narita}, M.~{Paegert},
  E.~{Palle}, F.~{Pepe}, J.~{Pepper}, A.~{Quirrenbach}, S.~A. {Rinehart},
  D.~{Sasselov}, B.~{Sato}, S.~{Seager}, A.~{Sozzetti}, K.~G. {Stassun},
  P.~{Sullivan}, A.~{Szentgyorgyi}, G.~{Torres}, S.~{Udry} and J.~{Villasenor},
  {Transiting Exoplanet Survey Satellite (TESS)}, {\em Journal of Astronomical
  Telescopes, Instruments, and Systems} {\bf 1}, p. 014003 (January 2015).

\bibitem{Rauer2014}
H.~{Rauer}, C.~{Catala}, C.~{Aerts}, T.~{Appourchaux}, W.~{Benz},
  A.~{Brandeker}, J.~{Christensen-Dalsgaard}, M.~{Deleuil}, L.~{Gizon}, M.~J.
  {Goupil}, M.~{G{\"u}del}, E.~{Janot-Pacheco}, M.~{Mas-Hesse}, I.~{Pagano},
  G.~{Piotto}, D.~{Pollacco}, {\.{C}}.~{Santos}, A.~{Smith}, J.~C.
  {Su{\'a}rez}, R.~{Szab{\'o}}, S.~{Udry}, V.~{Adibekyan}, Y.~{Alibert}, J.~M.
  {Almenara}, P.~{Amaro-Seoane}, M.~A.-v. {Eiff}, M.~{Asplund}, E.~{Antonello},
  S.~{Barnes}, F.~{Baudin}, K.~{Belkacem}, M.~{Bergemann}, G.~{Bihain}, A.~C.
  {Birch}, X.~{Bonfils}, I.~{Boisse}, A.~S. {Bonomo}, F.~{Borsa}, I.~M.
  {Brand{\~a}o}, E.~{Brocato}, S.~{Brun}, M.~{Burleigh}, R.~{Burston},
  J.~{Cabrera}, S.~{Cassisi}, W.~{Chaplin}, S.~{Charpinet}, C.~{Chiappini},
  R.~P. {Church}, S.~{Csizmadia}, M.~{Cunha}, M.~{Damasso}, M.~B. {Davies},
  H.~J. {Deeg}, R.~F. {D{\'\i}az}, S.~{Dreizler}, C.~{Dreyer},
  P.~{Eggenberger}, D.~{Ehrenreich}, P.~{Eigm{\"u}ller}, A.~{Erikson},
  R.~{Farmer}, S.~{Feltzing}, F.~{de Oliveira Fialho}, P.~{Figueira},
  T.~{Forveille}, M.~{Fridlund}, R.~A. {Garc{\'\i}a}, P.~{Giommi},
  G.~{Giuffrida}, M.~{Godolt}, J.~{Gomes da Silva}, T.~{Granzer}, J.~L.
  {Grenfell}, A.~{Grotsch-Noels}, E.~{G{\"u}nther}, C.~A. {Haswell}, A.~P.
  {Hatzes}, G.~{H{\'e}brard}, S.~{Hekker}, R.~{Helled}, K.~{Heng}, J.~M.
  {Jenkins}, A.~{Johansen}, M.~L. {Khodachenko}, K.~G. {Kislyakova}, W.~{Kley},
  U.~{Kolb}, N.~{Krivova}, F.~{Kupka}, H.~{Lammer}, A.~F. {Lanza},
  Y.~{Lebreton}, D.~{Magrin}, P.~{Marcos-Arenal}, P.~M. {Marrese}, J.~P.
  {Marques}, J.~{Martins}, S.~{Mathis}, S.~{Mathur}, S.~{Messina}, A.~{Miglio},
  J.~{Montalban}, M.~{Montalto}, M.~J.~P.~F.~G. {Monteiro}, H.~{Moradi},
  E.~{Moravveji}, C.~{Mordasini}, T.~{Morel}, A.~{Mortier}, V.~{Nascimbeni},
  R.~P. {Nelson}, M.~B. {Nielsen}, L.~{Noack}, A.~J. {Norton}, A.~{Ofir},
  M.~{Oshagh}, R.~M. {Ouazzani}, P.~{P{\'a}pics}, V.~C. {Parro}, P.~{Petit},
  B.~{Plez}, E.~{Poretti}, A.~{Quirrenbach}, R.~{Ragazzoni}, G.~{Raimondo},
  M.~{Rainer}, D.~R. {Reese}, R.~{Redmer}, S.~{Reffert}, B.~{Rojas-Ayala},
  I.~W. {Roxburgh}, S.~{Salmon}, A.~{Santerne}, J.~{Schneider}, J.~{Schou},
  S.~{Schuh}, H.~{Schunker}, A.~{Silva-Valio}, R.~{Silvotti}, I.~{Skillen},
  I.~{Snellen}, F.~{Sohl}, S.~G. {Sousa}, A.~{Sozzetti}, D.~{Stello}, K.~G.
  {Strassmeier}, M.~{{\v{S}}vanda}, G.~M. {Szab{\'o}}, A.~{Tkachenko},
  D.~{Valencia}, V.~{Van Grootel}, S.~D. {Vauclair}, P.~{Ventura}, F.~W.
  {Wagner}, N.~A. {Walton}, J.~{Weingrill}, S.~C. {Werner}, P.~J. {Wheatley}
  and K.~{Zwintz}, {The PLATO 2.0 mission}, {\em Experimental Astronomy} {\bf
  38}, 249 (November 2014).

\end{thebibliography}

\end{document}